\documentclass[aps,twocolumn,preprintnumbers,superscriptaddress,amsmath,amssymb,amsfonts]{revtex4-2}
\usepackage{graphicx}  % needed for figures
\usepackage{color}
\usepackage[colorlinks,bookmarks=true,citecolor=blue,linkcolor=blue,urlcolor=blue, breaklinks=true]{hyperref}
\usepackage{bm}
\usepackage{amsmath,amssymb}
\begin{document}

% Use the \preprint command to place your local institutional report
% number in the upper righthand corner of the title page in preprint mode.
% Multiple \preprint commands are allowed.
% Use the 'preprintnumbers' class option to override journal defaults
% to display numbers if necessary
%\preprint{}
%%%%%%%%%%%%%%%%%%%%%%%%%%%%%%%%%%%%%%%%%%%%%%%%%%%%%%%%%%%%
%Title of paper
\title{Tuning from intrinsic to disorder-dominated quantum states in a kagome antiferromagnet}
%%%%%%%%%%%%%%%%%%%%%%%%%%%%%%%%%%%%%%%%%%%%%%%%%%%%%%%%%%%%
% repeat the \author .. \affiliation  etc. as needed
% \email, \thanks, \homepage, \altaffiliation all apply to the current
% author. Explanatory text should go in the []'s, actual e-mail
% address or url should go in the {}'s for \email and \homepage.
% Please use the appropriate macro foreach each type of information

% \affiliation command applies to all authors since the last
% \affiliation command. The \affiliation command should follow the
% other information
% \affiliation can be followed by \email, \homepage, \thanks as well.

\author{Y. Tanaka}
\affiliation{Department of Applied Physics, The University of Tokyo, Tokyo 113-8656, Japan}
\author{R. Kumakura}
\affiliation{Department of Applied Physics, The University of Tokyo, Tokyo 113-8656, Japan}
\author{T. Yamaguchi}
\affiliation{Department of Applied Physics, The University of Tokyo, Tokyo 113-8656, Japan}
\author{M. Kato}
\affiliation{Department of Applied Physics, The University of Tokyo, Tokyo 113-8656, Japan}
\author{K. Totsuka}
\affiliation{Yukawa Institute for Theoretical Physics, Kyoto University, Kyoto 606-8502, Japan}
\author{B. Li}
\affiliation{Wuhan National High Magnetic Field Center and School of Physics, Huazhong University of Science and Technology, 430074 Wuhan, China}
\author{Y. Zhao}
\affiliation{Wuhan National High Magnetic Field Center and School of Physics, Huazhong University of Science and Technology, 430074 Wuhan, China}
\author{Y. Li}
\affiliation{Wuhan National High Magnetic Field Center and School of Physics, Huazhong University of Science and Technology, 430074 Wuhan, China}
\author{R. Grasset}
\affiliation{Laboratoire des Solides Irradi\'{e}s, CEA/DRF/IRAMIS, Ecole Polytechnique, CNRS, Institut Polytechnique de Paris, F-91128 Palaiseau, France}
\author{O. Cavani}
\affiliation{Laboratoire des Solides Irradi\'{e}s, CEA/DRF/IRAMIS, Ecole Polytechnique, CNRS, Institut Polytechnique de Paris, F-91128 Palaiseau, France}
\author{A. Alessi}
\affiliation{Laboratoire des Solides Irradi\'{e}s, CEA/DRF/IRAMIS, Ecole Polytechnique, CNRS, Institut Polytechnique de Paris, F-91128 Palaiseau, France}
\author{V. Sharma}
\affiliation{Institute for Solid State Physics, The University of Tokyo, Kashiwa, Chiba 277-8581, Japan}
\author{Y. Kohama}
\affiliation{Institute for Solid State Physics, The University of Tokyo, Kashiwa, Chiba 277-8581, Japan}
\author{S. Suetsugu}
\email{suetsugu@ap.t.u-tokyo.ac.jp}
\affiliation{Department of Applied Physics, The University of Tokyo, Tokyo 113-8656, Japan}
%Collaboration name if desired (requires use of superscriptaddress
%option in \documentclass). \noaffiliation is required (may also be
%used with the \author command).
%\collaboration can be followed by \email, \homepage, \thanks as well.
%\collaboration{}
%\noaffiliation

\date{\today}

% insert suggested keywords - APS authors don't need to do this
%\keywords{}

%\maketitle must follow title, authors, abstract, and keywords

% body of paper here - Use proper section commands
% References should be done using the \cite, \ref, and \label commands
\begin{abstract}
The spin-1/2 kagome Heisenberg antiferromagnet (AFM) provides a paradigmatic platform for realizing quantum spin-liquid (QSL) and magnetization-plateau states. 
Experimental progress, however, has been hindered by structural disorder, whose effects on these quantum states remain elusive. A key experimental challenge is therefore to connect intrinsic and disorder-dominated regimes through controlled disorder and identify emergent degrees of freedom associated with disorder. Here we introduce controlled point defects through electron irradiation into the spin-1/2 kagome AFM YCu$_3$(OH)$_{6.5}$Br$_{2.5}$ (YCOB) and track the resulting evolution of its low-energy magnetic states. High-field magnetic torque measurements reveal that the 1/9 and 1/3 magnetization plateaus are strongly suppressed by only $\sim$0.05\% irradiation-induced defects, demonstrating that these delicate quantum many-body states are extraordinarily fragile against dilute disorder. Moreover, the irradiation induces a sign reversal of the magnetic torque only at low temperatures and low fields, indicating the emergence of correlated defect-induced moments rather than isolated free spins. Remarkably, the characteristic scaling of the magnetic specific heat, absent in pristine YCOB, progressively emerges with increasing disorder. These findings establish dilute disorder as an effective tuning parameter that drives the kagome AFM from intrinsic quantum states toward a disorder-dominated regime. The contrasting responses further identify pristine YCOB as an exceptionally clean kagome AFM for accessing intrinsic quantum states.
\end{abstract}

\maketitle

\section{Introduction}
The spin-1/2 Heisenberg AFM on the two-dimensional (2D) kagome lattice has long served as a paradigmatic platform for exploring exotic quantum many-body states, owing to strong geometrical frustration and quantum fluctuations \cite{PhysRevB.45.12377,balents2010spin}. In particular, this system is widely regarded as one of the most promising candidates for realizing a QSL state, characterized by long-range quantum entanglement and emergent fractionalized spin excitations. Despite decades of intensive studies, however, understanding the nature of the kagome AFM has been a long-standing issue in quantum magnetism \cite{PhysRevLett.98.117205,yan2011spin,jiang2012identifying,PhysRevLett.109.067201,PhysRevB.87.060405,PhysRevX.7.031020,PhysRevB.100.155142}. This difficulty arises from the presence of numerous competing low-energy states that lie very close in energy, making a reliable numerical identification of the ground state exceptionally challenging. To date, several candidate ground states have been proposed, among which a gapless $U(1)$ Dirac QSL \cite{PhysRevLett.98.117205,PhysRevB.87.060405,PhysRevX.7.031020} and a gapped $Z_2$ QSL \cite{yan2011spin,jiang2012identifying,PhysRevLett.109.067201,PhysRevB.100.155142} are regarded as the leading possibilities. 

Another important unresolved issue surrounding the kagome AFM concerns the ground state in external fields. Of particular interest are field-induced spin-gapped phases manifested as magnetization plateaus at 1/9, 1/3, 5/9, and 7/9 of the saturated moment \cite{PhysRevLett.78.1984,nishimoto2013controlling,PhysRevB.93.060407,chen2018thermodynamics,PhysRevB.107.L220401}. The 1/3, 5/9, and 7/9 plateau states are generally understood as magnon-crystal states formed by localized hexagonal magnons, whose motion is suppressed by kinetic frustration inherent to the geometry of the kagome lattice. In contrast, the nature of the 1/9 plateau remains unsettled, with proposed candidates including $Z_3$ QSL \cite{nishimoto2013controlling,PhysRevLett.133.096501} and valence-bond-crystal states \cite{PhysRevB.93.060407,PhysRevB.107.L220401,morita2024valence}. 

Experimentally accessing these enigmatic QSL and magnetization-plateau states is therefore essential for resolving the intrinsic nature of the spin-1/2 kagome AFM. To date, the most extensively studied candidate material has been herbertsmithite ZnCu$_3$(OH)$_6$Cl$_2$ \cite{mendels2010quantum,shores2005structurally,PhysRevLett.98.077204,PhysRevLett.98.107204,han2012fractionalized,fu2015evidence,freedman2010site,PhysRevB.76.132411,PhysRevB.106.174406,PhysRevLett.104.147201} which features a 2D kagome lattice of Cu$^{2+}$ ions. Despite its dominant nearest-neighbor Heisenberg antiferromagnetic exchange of $J \sim 200$\,K, this compound exhibits no magnetic order down to 20\,mK. Inelastic neutron scattering measurements have reported a continuous scattering consistent with fractionalized spinon excitations \cite{han2012fractionalized}. However, herbertsmithite suffers from inevitable antisite disorder between Cu$^{2+}$ ions in the kagome layers and Zn$^{2+}$ ions between the layers \cite{shores2005structurally,freedman2010site}, with the disorder level reported to be as high as 5--15\% \cite{freedman2010site,fu2015evidence,PhysRevLett.104.147201,PhysRevB.76.132411,PhysRevB.106.174406}. This intersite mixing introduces orphan spins, which dominate low-energy thermodynamic responses, as exemplified by the characteristic data collapse of the low-temperature specific heat and magnetization \cite{PhysRevLett.98.107204,PhysRevLett.104.147201,PhysRevB.106.174406,kimchi2018scaling}. Moreover, the magnetization plateaus anticipated for the kagome AFM have not been discernibly observed even under extremely high magnetic fields \cite{han2014thermodynamic}. These limitations highlight the need for a cleaner kagome AFM in which the intrinsic QSL and magnetization-plateau states can be directly investigated. Such a material would also provide a starting point for tuning disorder, allowing us to follow the evolution of its low-energy magnetic response from intrinsic to disorder-dominated behavior.

Recently discovered YCu$_3$(OH)$_{6.5}$Br$_{2.5}$ (YCOB) \cite{chen2020quantum,PhysRevB.105.024418,PhysRevB.105.L121109,lu2022observation} has emerged as a promising platform for investigating the intrinsic ground states of the spin-1/2 kagome AFM [Figs.\,1(a) and 1(b)]. This compound consists of the 2D kagome layers of Cu$^{2+}$ ions and exhibits no magnetic order down to 50\,mK \cite{PhysRevB.105.L121109} despite sizable antiferromagnetic interactions of $J=$ 40--80\,K. A key advantage of YCOB is the absence of intersite mixing between magnetic Cu$^{2+}$ and nonmagnetic Y$^{3+}$ ions, owing to the large difference in their ionic radii. This preserves an intact Cu$^{2+}$ kagome lattice and strongly suppresses disorder effects associated with orphan spins. The recent observation of the 1/9 and 1/3 magnetization plateaus \cite{suetsugu2023emergent,jeon2024,zheng2023unconventional} demonstrates that YCOB can support delicate field-induced quantum phases that have been inaccessible in more disordered kagome AFMs.

Nevertheless, whether YCOB truly provides access to the clean spin-1/2 kagome AFM has remained highly controversial. X-ray diffraction measurements have suggested that random occupation of Br$^-$ and OH$^-$ ions displaces approximately 70\% of Y$^{3+}$ ions from their optimal positions \cite{PhysRevB.105.024418}. Such local structural distortions may modify the Cu--O--Cu bond angles and introduce randomness into the superexchange interactions. Indeed, such bond disorder has been used to interpret several experimental observations \cite{PhysRevB.105.024418,PhysRevB.109.104403,PhysRevB.110.064418}, and has also been discussed as a possible ingredient of the putative QSL state \cite{PhysRevB.110.085146}. If this bond randomness were strong enough to dominate the low-energy physics, it would obscure the intrinsic nature of the kagome AFM and fundamentally limit the significance of YCOB as a model material.

However, because such interpretations often rely on model-dependent analyses and the microscopic origin of the disorder-induced responses remains unclear, whether such severe bond randomness is actually present requires careful scrutiny. In fact, several observations call for a careful reassessment of the effects of disorder. For example, inelastic neutron scattering measurements have revealed sharp dispersive spin excitations rather than the strongly damped excitations expected for a highly disordered system \cite{zeng2023dirac,qypv-n4yg}. Moreover, the well-defined 1/9 and 1/3 magnetization plateaus observed in YCOB \cite{suetsugu2023emergent,jeon2024,zheng2023unconventional} suggest that the Cu$^{2+}$ kagome network remains sufficiently clean to stabilize delicate field-induced quantum states, since numerical studies have shown that strong bond randomness suppresses such plateau anomalies \cite{kawamura2014quantum,PhysRevB.92.134407}. Thus, it is premature to conclude that YCOB hosts bond randomness strong enough to obscure the intrinsic nature of the kagome AFM. A quantitative assessment of disorder effects in YCOB is therefore essential for clarifying both its intrinsic quantum states and the emergent degrees of freedom associated with disorder.

Here we investigate disorder effects in YCOB by artificially introducing point defects through electron irradiation [Fig.\,1(c)]. High-energy electron beams create Frenkel pairs consisting of vacancies and interstitial atoms, which act as point defects \cite{PhysRevX.14.011045}. Since the defect density can be systematically controlled by the irradiation fluence, this approach enables us to track how dilute disorder evolves the low-energy magnetic states. We find that even a minute concentration of defects strongly suppresses the 1/9 and 1/3 magnetization plateaus, demonstrating their extreme sensitivity to dilute disorder. Strikingly, the scaling behavior of the magnetic specific heat, characteristic of frustrated disordered quantum magnets, progressively emerges with irradiation fluence. The present results establish dilute disorder as an effective tuning parameter directly connecting intrinsic quantum states to a disorder-dominated regime in the kagome AFM.

\section{Methods}
\subsection{Single crystal growth and electron irradiation}
Single crystals of YCOB were grown by a hydrothermal method as reported previously \cite{PhysRevB.105.024418}. In this study, we investigated the same single crystals used in Ref.\,\citenum{suetsugu2023emergent}, which exhibit clear 1/9 and 1/3 magnetization plateaus. We examined pristine YCOB single crystals and crystals irradiated with electron fluences of 1\,C/cm$^2$ and 5\,C/cm$^2$. These samples are hereafter referred to as the pristine, 1\,C/cm$^2$, and 5\,C/cm$^2$ samples, respectively. Electron irradiation with the incident electron energy of 2.5\,MeV was performed on SIRIUS Pelletron accelerator operated by the Laboratoire des Solides irradi\'{e}s (LSI) at \'{E}cole Polytechnique. At this energy, electrons have a large penetration depth on the order of millimeters, ensuring a uniform distribution of point defects within the samples. While vacancies on the Cu sites are expected to act as spin vacancies, vacancies on the other atomic sites can introduce bond disorder by locally modifying the Cu--O--Cu bond angles. The irradiation was performed at a low temperature of $\sim$20\,K with the sample immersed in liquid hydrogen, after which the samples were warmed up to room temperature. Because the Frenkel pairs (vacancy-interstitial pairs) induced by irradiation can be partially annealed during the warming process, the defect density calculated from the scattering cross-section (see Fig.\,S1 in Supplemental Material \cite{supp}) represents the upper limit of the actual disorder concentration. As summarized in Table S1 in Supplemental Material \cite{supp}, the estimated concentrations of irradiation-induced vacancies are much smaller than 0.1\% per 1\,C/cm$^2$ irradiation fluence for all atomic sites, indicating that only a minute amount of disorder is introduced into the present irradiated samples. It should be noted that the upper limits of the vacancy concentrations at Cu sites calculated for the fluences of 1\,C/cm$^2$ and 5\,C/cm$^2$ are 0.052\% and 0.261\%, which are orders of magnitude smaller than the concentrations of Cu/Zn intersite mixing reported in herbertsmithite \cite{freedman2010site,fu2015evidence,PhysRevLett.104.147201,PhysRevB.76.132411,PhysRevB.106.174406}.

\subsection{Magnetic torque experiments}
Magnetic torque measurements were performed using a piezoresistive cantilever technique \cite{RSI.73.3022}. For the experiments up to 57\,T, magnetic fields were generated using a non-destructive pulse magnet installed at the International MegaGauss Science Laboratory of the Institute for Solid State Physics at the University of Tokyo. The crystallographic $c$-axis was first identified as the field direction at which the low-field torque vanishes and changes sign upon varying the field orientation. The high-field measurements were then conducted at small tilt angles $\theta$ of $5.7^\circ$--$13.0^\circ$ relative to the $c$-axis. The magnetic torque up to 14\,T was measured using a superconducting magnet.

\subsection{Specific heat and magnetization experiments}
Specific heat experiments were conducted by a long-relaxation time method \cite{PhysRevB.63.094508,PhysRevLett.99.057001} by applying field $H$ along the crystal $c$-axis up to 14\,T. The magnetic contribution $C_\mathrm{mag}$ was obtained by subtracting the phonon contribution, which was estimated by fitting the high-temperature data measured in zero magnetic field up to 60\,K using a Physical Property Measurement System (see Fig.\,S2 in Supplemental Material \cite{supp}). Magnetization experiments were conducted using a commercial SQUID magnetometer with magnetic fields up to 7\,T applied parallel and perpendicular to the $c$-axis.

\section{Results}
\subsection{Magnetic susceptibility under controlled disorder}
We first examine the magnetic susceptibility to assess how electron irradiation modifies the magnetic response of YCOB. The inset of Fig.\,1(d) displays the temperature dependence of the magnetic susceptibility $\chi(T)$ for all samples in the low-temperature region. In the pristine sample, $\chi(T)$ exhibits an upturn below approximately 30\,K, which can be attributed, at least in part, to a Curie-like contribution from impurity spins. This low-temperature upturn is progressively enhanced with increasing irradiation fluence. In contrast, the linear fits to $\chi^{-1}$ for $T >150$\,K yield similar values of the Curie-Weiss temperatures $-89$, $-93$, and $-84$\,K for pristine, 1\,C/cm$^2$, and 5\,C/cm$^2$ samples, respectively [main panel of Fig.\,1(d)]. These results demonstrate that electron irradiation introduces controlled disorder without significantly altering the strength of the dominant exchange interactions.

\subsection{Fragility of magnetization plateaus against dilute disorder}
To investigate the robustness of the magnetization plateaus, we performed magnetic torque measurements with a piezoresistive cantilever in pulsed fields up to 57\,T. The torque $\bm{\tau}$ is given by $\bm{\tau} = \mu_0 V \bm{M} \times \bm{H}$, where $\bm{M}$ is the magnetization, $\mu_0$ is vacuum permeability, and $V$ is the sample volume. Figure 2(a) shows the field dependence of the torque amplitude divided by field $\tau/H$, which approximately reflects the magnetization, at 1.5\,K for the pristine ($\theta = 5.7^\circ$), 1\,C/cm$^2$ ($\theta = 13.0^\circ$), and 5\,C/cm$^2$ ($\theta = 10.1^\circ$) samples. For the pristine sample, plateau-like anomalies are clearly observed around 20 and 50\,T, corresponding to the previously reported 1/9 and 1/3 magnetization plateaus \cite{suetsugu2023emergent,jeon2024,zheng2023unconventional}, respectively. These anomalies are more distinctly resolved in the field derivative $d(\tau/H)/dH$, where sharp reductions of approximately 80\% and nearly 100\% are observed for the 1/9 and 1/3 plateaus, respectively [Fig.\,2(b)]. At higher temperatures, these plateaus become less pronounced, consistent with thermal broadening [Fig.\,S3(a) in Supplemental Material \cite{supp}].

In stark contrast, these plateau-like behaviors are strongly suppressed in the irradiated samples, as shown in Fig.\,2 [see also Figs.\,S3(b) and S3(c) in Supplemental Material \cite{supp}]. In the 1/3 plateau region, the reduction in $d(\tau/H)/dH$ decreases from nearly 100\% in the pristine sample to approximately 20\% in the 1\,C/cm$^2$ sample, and becomes absent within experimental uncertainty in the 5\,C/cm$^2$ sample, even at the lowest measured temperature of $T=1.5$\,K. The 1/9 plateau is almost completely smeared out already in the 1\,C/cm$^2$ sample, corresponding to only $\sim$0.05\% point defects. These results demonstrate that the magnetization-plateau states in YCOB are extremely sensitive to disorder. It should be noted, however, that the onset field of the 1/3 plateau in the 1\,C/cm$^2$ sample remains nearly the same as that in the pristine sample, indicating that the strength of the dominant exchange interactions is not substantially altered by electron irradiation.

This strong disorder sensitivity of the plateau states can be naturally understood in terms of disorder-induced in-gap states inside the field-induced spin gap. 
In magnetization-plateau states, spin rotation is prohibited by the opening of a spin gap $\Delta$, giving rise to a finite field range over which the magnetization remains nearly constant. This phenomenon has long been discussed in analogy with the quantum Hall effect (QHE) \cite{PhysRevLett.78.1984}, and the 1/3 and 1/9 magnetization plateaus in the kagome AFM are reminiscent of the integer and fractional QHEs, respectively. Disorder locally modifies the exchange interactions and creates regions in which the energy cost of magnetic excitations is lower than the bulk spin gap. Magnetic excitations can therefore nucleate around defects, producing in-gap states within the field-induced spin gap $\Delta$, as schematically shown in the inset of Fig. 2(a). Such in-gap states allow spin rotation to occur even at fields inside the plateau region, smearing out the magnetization plateau. The more rapid suppression of the 1/9 plateau than the 1/3 plateau is consistent with the smaller spin gap protecting the 1/9 plateau. Although the 1/9 plateau is more rapidly suppressed, both the 1/9 and 1/3 plateaus are strongly affected by only minute concentrations of irradiation-induced defects. This pronounced response demonstrates that delicate quantum many-body states in the kagome AFM are extraordinarily fragile against dilute disorder and provides a high-field manifestation of disorder tuning. Conversely, their clear observation in the pristine sample strongly indicates that pristine YCOB is an exceptionally clean kagome AFM with minimal disorder effects.

\subsection{Anisotropic response of defect-induced moments}
We further examine the low-field torque response, where disorder-induced contributions are expected to be most pronounced. In the pristine sample, $\tau/H$ increases monotonically with $H$ and remains positive in the measured field range. In contrast, the irradiated samples exhibit a distinct sign reversal of $\tau/H$, taking negative values in the low-field region and positive values at higher fields. As shown in Fig.\,2(a) (see also Fig.\,S4 in Supplemental Material \cite{supp}), this sign reversal becomes more pronounced with increasing irradiation fluence. The temperature dependence measured up to 14\,T further reveals that the sign reversal appears only in the low-temperature and low-field region, where the disorder-induced Curie-like contribution becomes dominant [Figs.\,3(a) and (b)]. Since an isotropic free-spin contribution does not generate magnetic torque, the observed sign reversal demonstrates that these disorder-induced moments possess an anisotropic magnetic response. Moreover, its sign is opposite to that of the intrinsic contribution from the kagome plane. Together with the scaling behavior discussed below, these results suggest that the disorder-induced moments are interacting or correlated rather than simple isolated free spins. It should be noted that a similar sign reversal has been reported in herbertsmithite \cite{PhysRevB.90.064417}, where orphan spins produce a complex field and temperature evolution of the torque response. The contrasting behavior between the pristine YCOB sample and the irradiated samples, as well as herbertsmithite, further supports the minimal disorder effects in pristine YCOB.

\subsection{Nonmonotonic evolution of the low-temperature specific heat}
We next examine how controlled disorder modifies the low-energy excitations through specific heat $C$ measurements. Figure 4(a) shows the temperature dependence of $C/T$ for all samples in zero field. The results for the pristine sample are in good agreement with previous reports \cite{PhysRevB.105.024418,PhysRevB.105.L121109,suetsugu2024gapless}, where the pronounced peak around 1.3\,K is consistent with numerical results for the kagome Heisenberg AFM \cite{PhysRevB.98.094423}. In the 1\,C/cm$^2$ sample, the peak magnitude of $C/T$ is reduced to approximately half of that in the pristine sample, indicating that the magnetic specific heat is strongly suppressed by only $\sim$0.05\% defects. Upon further irradiation to 5\,C/cm$^2$, however, the magnitude of $C$ increases again. This nonmonotonic behavior suggests a crossover from intrinsic spin excitations associated with the kagome planes, which are strongly suppressed by weak disorder, to a disorder-induced contribution that becomes dominant at higher irradiation fluence. This strong disorder sensitivity of the magnetic specific heat underscores the importance of clean materials for accessing the intrinsic spin excitations inherent to the kagome planes.

To disentangle the intrinsic contribution from the disorder-induced contribution, we further investigate $C/T$ under magnetic fields $\bm{H}$ parallel to the $c$-axis for the pristine and 5\,C/cm$^2$ samples [Figs.\,4(b) and (c)]. The most prominent feature is the contrasting field evolution of the peak observed at $\sim 1$\,K in zero field. In the pristine sample, the peak is suppressed with increasing $H$, while the peak temperature remains nearly independent of $H$. In contrast, in the 5\,C/cm$^2$ sample, the peak is not only suppressed but also shifts systematically to higher temperatures with increasing $H$. To track this field evolution more clearly, we determine the peak temperature $T_{\mathrm{peak}}$ from the peak position in $[C(H)-C(14\ \mathrm{T})]/T$, where the 14\,T data are subtracted as a background (Fig.\,S5 in Supplemental Material \cite{supp}). This analysis shows that $T_{\mathrm{peak}}$ for the pristine sample is almost field independent, whereas $T_{\mathrm{peak}}$ for the 1\,C/cm$^2$ and 5\,C/cm$^2$ samples increases roughly linearly with $H$. The nearly linear field dependence of $T_{\mathrm{peak}}$ in the irradiated samples suggests that the low-temperature specific heat is governed by a scaling form involving $T/H$. This behavior is reminiscent of the $T/H$ scaling and associated data collapse observed in various frustrated disordered quantum magnets including herbertsmithite \cite{PhysRevLett.98.107204,PhysRevB.106.174406,kitagawa2018spin}, where such scaling behaviors have been interpreted in terms of random-singlet physics \cite{kimchi2018scaling}.

\subsection{Emergence of thermodynamic scaling with disorder}
In frustrated quantum spin systems with substantial disorder, the magnetic specific heat $C_\mathrm{mag}$ and magnetization $M$ have been proposed to exhibit characteristic scaling \cite{kimchi2018scaling}. In this picture, the majority of spin-1/2 moments form a quantum paramagnetic state, while the remaining spins constitute a random network of singlets with a broad distribution of exchange interactions. The thermodynamic response is then dominated by this random-singlet subsystem, leading to scaling behavior in $(\mu_0 H)^\gamma C_\mathrm{mag} / T$ as a function of $T/H$ and in $M T^{\gamma - 1}$ as a function of $H/T$. Here $\gamma$ ($0 < \gamma < 1$) characterizes the power-law distribution of exchange interactions and serves as an effective measure of disorder within this description.

Figures\,5(a)--(c) show the scaling plots of the magnetic specific heat for the pristine, 1\,C/cm$^2$, and 5\,C/cm$^2$ samples, respectively, while Figs.\,5(d)--(f) show the corresponding scaling plots of the magnetization. Scaling plots of the specific heat without phonon subtraction are shown in Fig.\,S6 of the Supplemental Material \cite{supp}. The scaling exponent $\gamma$ was optimized for each sample to obtain the best overall collapse of both the magnetic specific heat and magnetization data, yielding $\gamma =$ 0.16, 0.37, and 0.53 for the pristine, 1\,C/cm$^2$, and 5\,C/cm$^2$ samples, respectively. The magnetization data exhibit good scaling collapse for all samples. However, for the pristine sample, the optimized value of $\gamma$ is close to zero, in which case the apparent scaling relation essentially reduces to the conventional linear response $M \propto H$. Thus, the apparent magnetization scaling does not distinguish random-singlet behavior from the conventional linear response. Indeed, no corresponding scaling collapse is observed in the specific heat, as expected from the nearly field-independent $T_{\mathrm{peak}}$ which shifts the peak position $T_\mathrm{peak}/H$ with field. The thermodynamic response of the pristine sample is therefore incompatible with the proposed random-singlet scaling form. In sharp contrast, the specific heat of the 5\,C/cm$^2$ sample exhibits a reasonable data collapse, while the 1\,C/cm$^2$ sample shows intermediate behavior between the pristine and 5\,C/cm$^2$ samples. Furthermore, the optimized $\gamma$ value increases monotonically with irradiation fluence. As shown in the inset of Fig.\,5(a), the increase in $\gamma$ accompanies the progressive suppression of the 1/3 plateau, suggesting the scaling response systematically evolves by the controlled introduction of disorder. The development of data collapse of the magnetic specific heat, together with the disorder-induced sign reversal of the magnetic torque, indicates the emergence of a disorder-dominated regime characterized by correlated defect-induced moments. These results establish a direct experimental connection between intrinsic quantum states and a disorder-dominated regime exhibiting the characteristic scaling.

\section{Discussion}
The combined torque and specific heat results reveal two complementary aspects of disorder tuning in a kagome AFM. Dilute disorder strongly perturbs the intrinsic response, as demonstrated by the suppression of the magnetization plateaus and the low-temperature specific heat. With increasing irradiation fluence, an additional magnetic contribution becomes prominent, accompanied by the sign reversal of the torque and the development of specific heat scaling. This evolution indicates that correlated defect-induced local moments become important low-energy degrees of freedom and drive the measured response toward a disorder-dominated regime. The nearly unchanged Curie-Weiss temperatures and onset field of the 1/3 plateau indicate that this disorder tuning occurs without a substantial change in the dominant exchange energy scale.

Despite the reasonable data collapse observed for the specific heat of the 5\,C/cm$^2$ sample, its detailed behavior deviates from the theoretical predictions \cite{kimchi2018scaling}. In the low-field limit, the scaling form requires $C_\mathrm{mag}/T \propto T^{-\gamma}$. In contrast, the observed $C_\mathrm{mag}/T$ decreases toward the lowest temperatures in zero and low fields rather than exhibiting the divergence. Moreover, in the low-temperature and high-field limit $T/H\ll 1$, the collapsed data approximately follow $(T/H)^{0.5}$, rather than either of the proposed integer power laws $(T/H)^q$ with $q=0$, 1, or 2 \cite{kimchi2018scaling}. These discrepancies may arise from limitations of the dimer approximation, since singlets involving multiple spins may modify the effective power laws. Alternatively, the correlated defect-induced moments generated by spin vacancies and bond randomness may form a state that differs microscopically from the canonical random-singlet scenario. We therefore interpret the irradiated samples in terms of a regime of correlated defect-induced moments rather than identify a specific random-singlet ground state. The emergence of data collapse despite these quantitative deviations suggests that the scaling phenomenon can persist beyond the specific microscopic assumptions and power laws of the existing theory.

Previous studies have discussed Br/OH structural randomness in YCOB as a possible origin of deviations from the ideal nearest-neighbor kagome Heisenberg model \cite{PhysRevB.105.024418,PhysRevB.109.104403,PhysRevB.110.064418}, and in some cases as an ingredient of the putative QSL state \cite{PhysRevB.110.085146}. Our controlled-disorder results provide a direct experimental constraint on this issue. Electron irradiation introduces structural defects that generate spin vacancies and bond randomness, enabling us to examine how sensitive the putative QSL state and magnetization-plateau states of YCOB are to disorder. The magnetization plateaus and intrinsic magnetic specific heat are strongly affected by only a minute concentration of irradiation-induced defects, demonstrating that these quantum states are highly sensitive to disorder. These findings therefore indicate that any intrinsic structural randomness in YCOB, if present, cannot act as a dominant perturbation that destroys or strongly obscures the QSL and magnetization-plateau states. A more plausible origin of the remaining deviations from the ideal nearest-neighbor kagome Heisenberg model is the presence of additional magnetic interactions beyond the nearest-neighbor Heisenberg exchange, as suggested by the magnetic anisotropy between the in-plane and out-of-plane directions \cite{PhysRevB.105.024418} and the finite slopes of the magnetization plateaus \cite{suetsugu2023emergent}. 

In conclusion, we have used electron irradiation to introduce controlled point defects and systematically tune the low-energy magnetic response of YCOB from intrinsic behavior toward a disorder-dominated regime. The rapid suppression of the magnetization plateaus and intrinsic low-temperature specific heat demonstrates the extraordinary fragility of quantum many-body states against dilute disorder. The progressive evolution of specific heat scaling, together with the sign reversal of magnetic torque, reveals the emergence of a disorder-dominated regime characterized by correlated defect-induced moments. These contrasting responses between pristine and irradiated YCOB identify pristine YCOB as an exceptionally clean kagome AFM and establish controlled disorder as an effective tuning parameter for frustrated quantum magnetism.

\begin{acknowledgments}
We thank T. Asaba, Y. Matsuda, C. Hotta, T. Misawa, T. Shimokawa, M. G. Yamada, K. Ishihara, and T. Shibauchi for insightful discussions. This work is supported by Grants-in-Aid for Scientific Research (KAKENHI) (No. JP25K00958), and Transformative Research Areas A “Extreme Universe” (No. 24H00965) from the Japan Society for the Promotion of Science. The work in Wuhan was supported by the National Natural Science Foundation of China (No. 12274153). The authors acknowledge support from the EMIR\&A French Network (FR CNRS 3618) on the platform SIRIUS.
\end{acknowledgments}

% \bibliography{ref.bib}
%

\begin{figure}
	\includegraphics[clip,width=8.5cm]{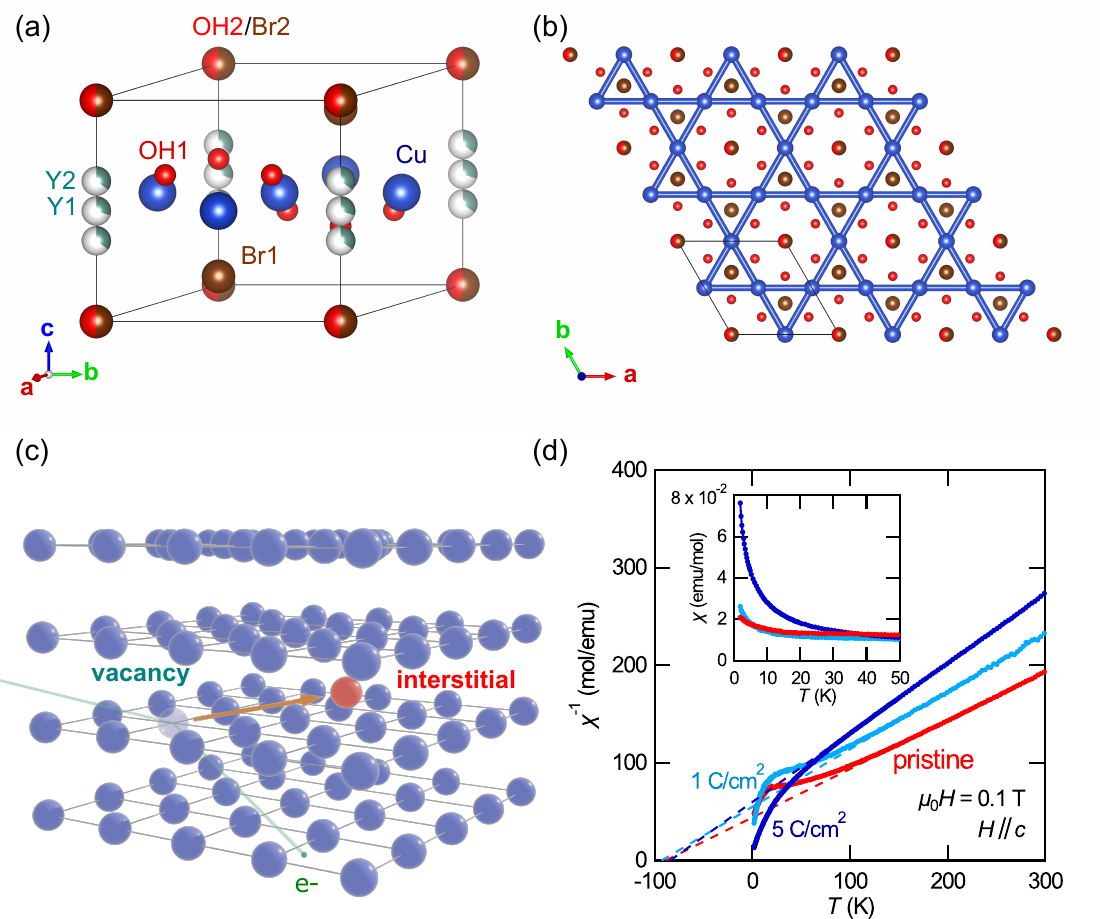}
	\caption{Crystal structure, electron irradiation, and magnetic susceptibility of YCOB. (a) Crystal structure of YCOB. X-ray diffraction measurements have suggested that random occupation of OH2 and Br2 sites displaces approximately 70\% of Y$^{3+}$ ions from their optimal positions on the kagome plane \cite{PhysRevB.105.024418}. (b) Kagome plane formed by Cu$^{2+}$ ions in YCOB. Owing to the large difference in ionic radii between magnetic Cu$^{2+}$ and nonmagnetic Y$^{3+}$ ions, YCOB preserves an intact Cu$^{2+}$ kagome lattice. (c) Schematic illustration of electron irradiation. High-energy electron beams create Frenkel pairs consisting of vacancies and interstitial atoms, which act as point defects. (d) Temperature dependence of the inverse magnetic susceptibility $\chi^{-1}(T)$ for the pristine, 1\,C/cm$^2$, and 5\,C/cm$^2$ samples measured at 0.1\,T for $\bm{H}\parallel c$. The dashed lines represent linear fits for $T>150$\,K, yielding similar values of the Curie-Weiss temperatures $-89$, $-93$, and $-84$\,K for pristine, 1\,C/cm$^2$, and 5\,C/cm$^2$ samples, respectively. The inset shows temperature dependence of $\chi(T)$ for the pristine, 1\,C/cm$^2$, and 5\,C/cm$^2$ samples in the low-temperature region. The low-temperature upturn arising from a Curie-like contribution is progressively enhanced with increasing irradiation fluence, indicating that electron irradiation introduces disorder into YCOB.
	\label{fig:crystal}
	}
\end{figure}

\begin{figure}
	\includegraphics[clip,width=8.5cm]{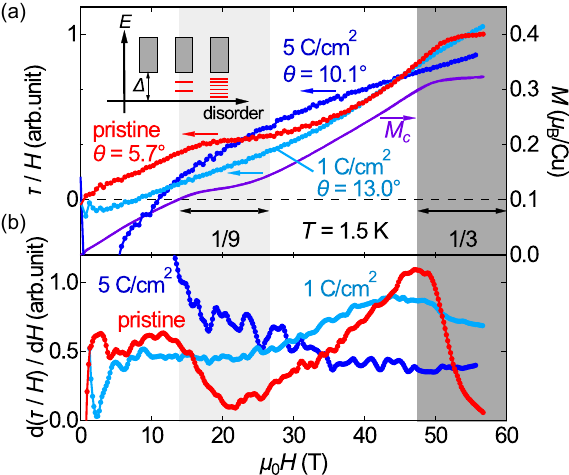}
	\caption{Magnetic torque and disorder fragility of magnetization plateaus in YCOB. (a) Field dependence of the magnetic torque divided by field $\tau/H$ at 1.5\,K for the pristine ($\theta = 5.7^\circ$), 1\,C/cm$^2$ ($\theta = 13.0^\circ$), and 5\,C/cm$^2$ ($\theta = 10.1^\circ$) samples. For comparison, the field dependence of the magnetization $M$ measured at 1.6\,K for $\bm{H}||c$ is also shown \cite{suetsugu2023emergent}. For the pristine sample, the plateau-like anomalies around 20 and 50\,T correspond to the 1/9 and 1/3 plateaus, as highlighted by light and dark gray regions, respectively. These anomalies are strongly suppressed in the irradiated samples, demonstrating that the magnetization-plateau states in YCOB are extremely sensitive to disorder. The inset schematically illustrates the effect of disorder on the field-induced spin gap $\Delta$ associated with a magnetization plateau. Disorder-induced in-gap (red lines) states allow spin flips to occur within the plateau region, smearing out the plateau. (b) Field dependence of the derivative $d(\tau/H)/dH$ at 1.5\,K. The sharp reductions in the plateau regions for the pristine sample become progressively suppressed with increasing irradiation fluence.
	\label{fig:torque} 
	}
\end{figure}

\begin{figure}
	\includegraphics[clip,width=8.5cm]{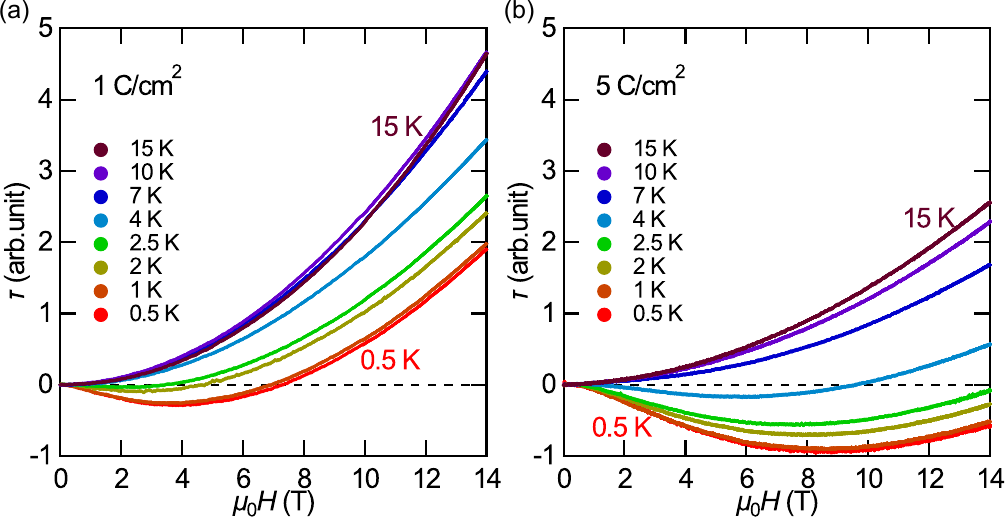}
	\caption{Magnetic torque of irradiated YCOB at low fields. (a, b) Field dependence of the magnetic torque $\tau$ for 1\,C/cm$^2$ (a) and 5\,C/cm$^2$ (b) samples, respectively, measured at $\theta = 10.0^\circ$ at several temperatures. The distinct sign reversal of $\tau$ at low fields becomes more pronounced with increasing irradiation fluence and decreasing temperature.
	\label{fig:torque_14T}
	}
\end{figure}

\begin{figure*}
	\includegraphics[clip,width=16cm]{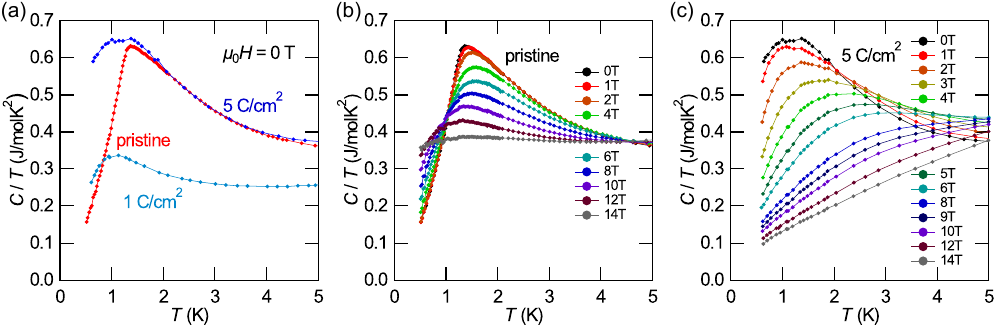}
	\caption{Specific heat of pristine and irradiated YCOB. (a) Temperature dependence of the specific heat divided by temperature $C/T$ in zero field for the pristine, 1\,C/cm$^2$, and 5\,C/cm$^2$ samples. The peak observed in the pristine sample is initially suppressed by weak disorder in the 1\,C/cm$^2$ sample, whereas stronger disorder in the 5\,C/cm$^2$ sample gives rise to a dominant contribution arising from disorder. (b, c) Temperature dependence of $C/T$ measured under several magnetic fields for the pristine (b) and 5\,C/cm$^2$ (c) samples. The peaks around 1\,K in both samples are suppressed with increasing $H$. While the peak temperature $T_\mathrm{peak}$ remains nearly constant in the pristine sample, $T_\mathrm{peak}$ increases almost linearly with $H$ in the 5\,C/cm$^2$ sample (see also Fig.\,S5).
	\label{fig:C}
	}
\end{figure*}

\begin{figure*}
	\includegraphics[clip,width=15cm]{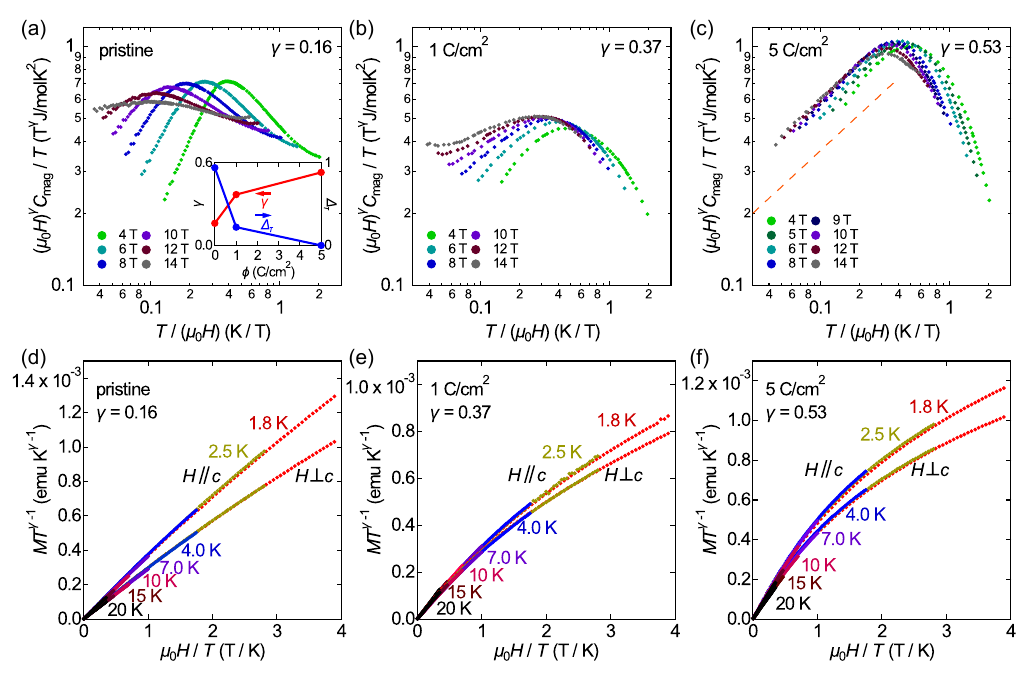}
	\caption{Scaling analysis of the magnetic specific heat and magnetization in pristine and irradiated YCOB. (a-c) Scaling plots of the magnetic specific heat, $(\mu_0 H)^\gamma C_\mathrm{mag}/T$, as a function of $T/H$ for the pristine (a), 1\,C/cm$^2$ (b), and 5\,C/cm$^2$ (c) samples. While the pristine sample shows no clear $T/H$ collapse, the 5\,C/cm$^2$ sample exhibits a reasonable data collapse. The 1\,C/cm$^2$ sample shows intermediate behavior between the pristine and 5\,C/cm$^2$ samples. For the 5\,C/cm$^2$ sample, the collapsed data approximately follow $(T/H)^{0.5}$ (dashed line) for $T/H \ll 1$. The inset in (a) displays the irradiation fluence dependence of $\gamma$ (left axis) and $\Delta_\tau$ (right axis), where $\Delta_\tau$ is the reduction ratio of $d(\tau/H)/dH$ in the 1/3 plateau region. (d-f) Scaling plots of the magnetization, $MT^{\gamma - 1}$, as a function of $H/T$ for the pristine (d), 1\,C/cm$^2$ (e), and 5\,C/cm$^2$ (f) samples. The data for both $\bm{H}||c$ and $\bm{H}\perp c$ are plotted. The scaling exponent $\gamma$ was optimized for each sample to obtain the best data collapse for both the magnetic specific heat and magnetization data.
	\label{fig:scaling}
	}
\end{figure*}

\end{document}

%%%%%%%%%%%%%%%%%%%%%%%%%%%%%%%%%%%%%%%%%%%%%%%%%
% Use the \preprint command to place your local institutional report
% number in the upper righthand corner of the title page in preprint mode.
% Multiple \preprint commands are allowed.
% Use the 'preprintnumbers' class option to override journal defaults
% to display numbers if necessary
%\preprint{}

%Title of paper
\title{Supplemental Material for Tuning from intrinsic to disorder-dominated quantum states in a kagome antiferromagnet}
% repeat the \author .. \affiliation  etc. as needed
% \email, \thanks, \homepage, \altaffiliation all apply to the current
% author. Explanatory text should go in the []'s, actual e-mail
% address or url should go in the {}'s for \email and \homepage.
% Please use the appropriate macro foreach each type of information

% \affiliation command applies to all authors since the last
% \affiliation command. The \affiliation command should follow the
% other information
% \affiliation can be followed by \email, \homepage, \thanks as well.
\author{Y. Tanaka}
\affiliation{Department of Applied Physics, The University of Tokyo, Tokyo 113-8656, Japan}
\author{R. Kumakura}
\affiliation{Department of Applied Physics, The University of Tokyo, Tokyo 113-8656, Japan}
\author{T. Yamaguchi}
\affiliation{Department of Applied Physics, The University of Tokyo, Tokyo 113-8656, Japan}
\author{M. Kato}
\affiliation{Department of Applied Physics, The University of Tokyo, Tokyo 113-8656, Japan}
\author{K. Totsuka}
\affiliation{Yukawa Institute for Theoretical Physics, Kyoto University, Kyoto 606-8502, Japan}
\author{B. Li}
\affiliation{Wuhan National High Magnetic Field Center and School of Physics, Huazhong University of Science and Technology, 430074 Wuhan, China}
\author{Y. Zhao}
\affiliation{Wuhan National High Magnetic Field Center and School of Physics, Huazhong University of Science and Technology, 430074 Wuhan, China}
\author{Y. Li}
\affiliation{Wuhan National High Magnetic Field Center and School of Physics, Huazhong University of Science and Technology, 430074 Wuhan, China}
\author{R. Grasset}
\affiliation{Laboratoire des Solides Irradi\'{e}s, CEA/DRF/IRAMIS, Ecole Polytechnique, CNRS, Institut Polytechnique de Paris, F-91128 Palaiseau, France}
\author{O. Cavani}
\affiliation{Laboratoire des Solides Irradi\'{e}s, CEA/DRF/IRAMIS, Ecole Polytechnique, CNRS, Institut Polytechnique de Paris, F-91128 Palaiseau, France}
\author{A. Alessi}
\affiliation{Laboratoire des Solides Irradi\'{e}s, CEA/DRF/IRAMIS, Ecole Polytechnique, CNRS, Institut Polytechnique de Paris, F-91128 Palaiseau, France}
\author{V. Sharma}
\affiliation{Institute for Solid State Physics, The University of Tokyo, Kashiwa, Chiba 277-8581, Japan}
\author{Y. Kohama}
\affiliation{Institute for Solid State Physics, The University of Tokyo, Kashiwa, Chiba 277-8581, Japan}
\author{S. Suetsugu}
\affiliation{Department of Applied Physics, The University of Tokyo, Tokyo 113-8656, Japan}
%Collaboration name if desired (requires use of superscriptaddress
%option in \documentclass). \noaffiliation is required (may also be
%used with the \author command).
%\collaboration can be followed by \email, \homepage, \thanks as well.
%\collaboration{}
%\noaffiliation
%%%%%%%%%%%%%%%%%%%%%%%%%%%%%%%%%%%%%%%%%%%%%%%%%
\date{\today}
%%%%%%%%%%%%%%%%%%%%%%%%%%%%%%%%%%%%%%%%%%%%%%%%%
% \begin{abstract}
% \end{abstract}

% insert suggested keywords - APS authors don't need to do this
%\keywords{}

%\maketitle must follow title, authors, abstract, and keywords
\maketitle
%%%%%%%%%%%%%%%%%%%%%%%%%%%%%%%%%%%%%%%%%%%%%%%%%
% body of paper here - Use proper section commands
%% References should be done using the \cite, \ref, and \label commands
%
\clearpage

\section{Estimation of irradiation-induced vacancy concentrations}

We estimated the concentrations of vacancies introduced by electron irradiation from the electron scattering cross sections for Frenkel-pair production. The cross sections $\sigma_i$ for each atomic species $i$ were calculated as a function of the incident electron energy using the SECTE software developed at Laboratoire des Solides Irradi\'{e}s \cite{bois1987}. In this calculation, the threshold displacement energy was taken to be a typical value of $E_{\mathrm{d}} = 25$\,eV for all atomic species \cite{roppongi2023bulk}. 

The vacancy density $n_i$ for atomic species $i$ is estimated as
\begin{equation}
n_i = \sigma_i \phi N_i / V,
\end{equation}
where $\phi$ is the irradiation fluence, $N_i$ is the number of atoms of species $i$ in the unit cell, and $V$ is the unit-cell volume. The vacancy concentration on sites of atomic species $i$ is therefore given by
\begin{equation}
c_i = \frac{n_i}{N_i/V} = \sigma_i \phi .
\end{equation}

Figure\,S1 shows the calculated cross sections $\sigma_i$ for Y, Cu, O, H, and Br sites as a function of incident electron energy $E$. Using the values of $\sigma_i$ at the incident electron energy of 2.5\,MeV, we estimated the vacancy concentrations in YCu$_3$(OH)$_{6.5}$Br$_{2.5}$ (YCOB) for irradiation fluences of 1 and 5\,C/cm$^2$, as summarized in Table\,S1. The estimated vacancy concentrations are well below 0.1\% per 1\,C/cm$^2$ irradiation fluence for all atomic species. Vacancies on Cu sites are expected to act as spin vacancies, whereas vacancies on the other atomic sites can introduce bond disorder by locally modifying the Cu--O--Cu bond angles.

We note that these estimates provide upper limits for the actual vacancy concentrations. The irradiation was performed at $\sim 20$\,K, and the samples were subsequently warmed to room temperature. Since part of the Frenkel pairs can be annealed during this warming process, the actual concentration of residual defects is expected to be smaller than the values listed in Table\,S1.

\clearpage

\begin{figure*}[h]
	\includegraphics[clip,width=8cm]{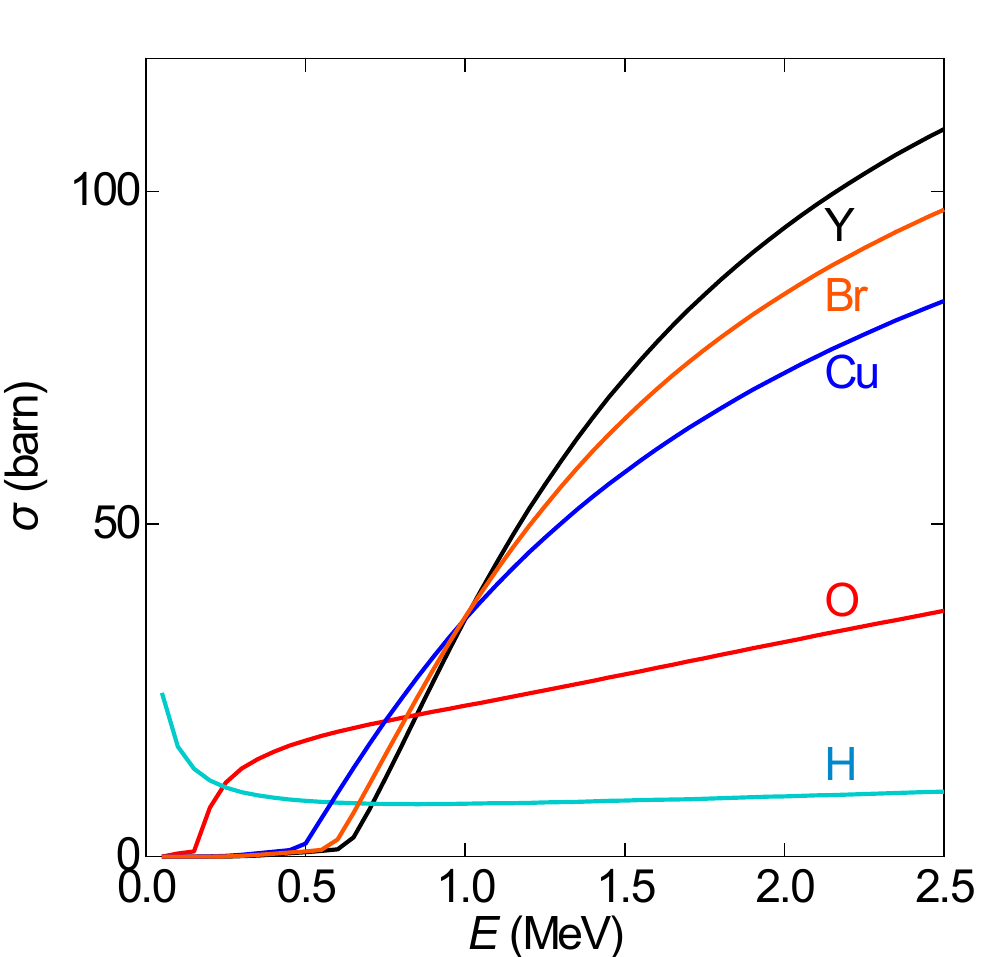}
	\caption{Calculated electron scattering cross sections $\sigma_i$ for Frenkel-pair production at Y, Cu, O, H, and Br sites in YCOB as a function of incident electron energy $E$.
	}
	\label{fig:scattering_cross_section}
\end{figure*}

\begin{table}
\centering
\caption{{\bf Estimated vacancy concentrations induced by electron irradiation.} The values are given as percentages of the corresponding atomic sites in YCOB for irradiation fluences of 1 and 5\,C/cm$^2$.}
\label{tab}
\begin{ruledtabular}
\begin{tabular}{lccccc}
Atomic site & Y & Cu & O & H & Br \\
\hline
1 C/cm$^2$ & 0.068\% & 0.052\% & 0.023\% & 0.006\% & 0.061\% \\
5 C/cm$^2$ & 0.341\% & 0.261\% & 0.115\% & 0.031\% & 0.304\% \\
\end{tabular}
\end{ruledtabular}
\end{table}

\clearpage

\section{Estimation of the phonon contribution to the specific heat}
To perform the scaling analysis presented in Fig.\,5 of the main text, we estimated the phonon contribution $C_{\mathrm{ph}}$ for the pristine, 1\,C/cm$^2$, and 5\,C/cm$^2$ samples and obtained the magnetic specific heat as $C_{\mathrm{mag}}(T,H)=C(T,H)-C_{\mathrm{ph}}(T)$. For the pristine sample, we used the phonon contribution reported in the literature \cite{PhysRevB.105.024418}. For the 1\,C/cm$^2$ and 5\,C/cm$^2$ samples, the phonon contributions were determined by fitting their zero-field specific-heat data using the Debye--Einstein phonon model employed in Ref.\,\citenum{PhysRevB.105.024418}. Figure\,S2 shows the zero-field specific-heat data and the corresponding fits for the 1\,C/cm$^2$ and 5\,C/cm$^2$ samples. The phonon contributions obtained from these fits were extrapolated to the low-temperature region and subtracted from the data measured at all magnetic fields. The resulting $C_{\mathrm{mag}}$ values were used for the scaling plots of $(\mu_0 H)^\gamma C_{\mathrm{mag}}/T$ as a function of $T/H$ shown in Fig.\,5 in the main text.

\begin{figure*}[h]
	\includegraphics[clip,width=12cm]{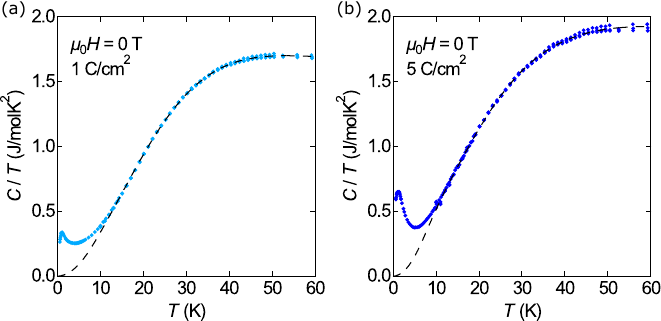}
	\caption{Estimation of the phonon contribution to the specific heat. (a, b) Temperature dependence of the zero-field specific heat divided by temperature of the 1\,C/cm$^2$ (a) and 5\,C/cm$^2$ (b) samples. The black dashed lines represent fits to the experimental data following the procedure used in the previous study \cite{PhysRevB.105.024418}.}
	\label{fig:phonon_fit}
\end{figure*}

\clearpage

\section{Temperature evolution of the magnetization plateaus}
Figure\,S3 shows the field dependence of the high-field magnetic torque divided by field $\tau/H$ at selected temperatures for the pristine and irradiated samples. For the pristine sample, the 1/9 and 1/3 plateaus, which are clearly observed at 1.5\,K, become less pronounced with increasing temperature [Fig.\,S3(a)]. For the irradiated samples, both plateaus are strongly suppressed even at the lowest measured temperature of 1.5\,K [Figs.\,S3(b) and S3(c)].

\begin{figure*}[h]
	\includegraphics[clip,width=16cm]{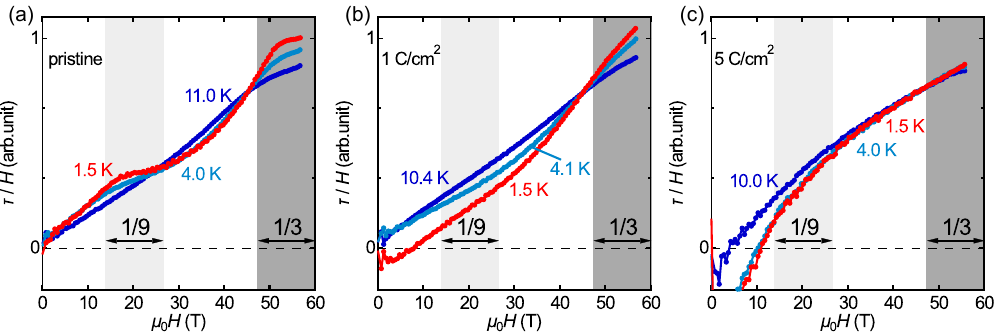}
	\caption{Field dependence of $\tau/H$ of pristine and irradiated YCOB up to 57\,T. (a-c) Field dependence of $\tau/H$ at selected temperatures for the pristine ($\theta = 5.7^\circ$), 1\,C/cm$^2$ ($\theta = 13.0^\circ$), and 5\,C/cm$^2$ ($\theta = 10.1^\circ$) samples, respectively. 
	\label{fig:tauH_temp_dep}
	}
\end{figure*}

\section{Disorder-induced torque response at low fields}
Here we examine the low-field torque response, where disorder-induced contributions are expected to be most pronounced. Figure S4 compares the pristine and irradiated samples in the low field region and at 1.5\,K. While $\tau$ monotonically increases with $H$ for the pristine sample, $\tau$ takes negative values in the low-field region and positive values at higher fields for irradiated samples. 
% Figure\,S5 shows the temperature evolution of $\tau$ for the irradiated samples. The sign reversal appears only in the low-temperature and low-field region and becomes more pronounced with increasing irradiation fluence.

\begin{figure*}[h]
	\includegraphics[clip,width=6cm]{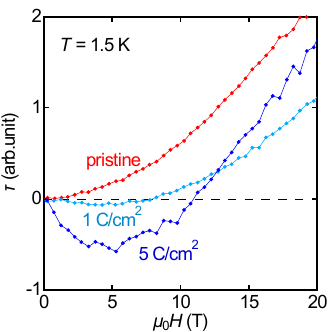}
	\caption{Field dependence of magnetic torque of pristine and irradiated YCOB in the low field region. Field dependence of $\tau$ at 1.5\,K up to 20\,T for the pristine ($\theta = 5.7^\circ$), 1\,C/cm$^2$ ($\theta = 13.0^\circ$), and 5\,C/cm$^2$ ($\theta = 10.1^\circ$) samples, respectively. 
	\label{fig:tau_low_temp_dose_dep}
	}
\end{figure*}

% \begin{figure*}[h]
% 	\includegraphics[clip,width=12cm]{extended_fig_torque_14T.pdf}
% 	\caption{Magnetic torque of irradiated YCOB at low fields. (a, b) Field dependence of the magnetic torque $\tau$ for 1\,C/cm$^2$ (a) and 5\,C/cm$^2$ (b) samples, respectively, measured at $\theta = 10.0^\circ$ at several temperatures. The distinct sign reversal of $\tau$ becomes more pronounced with increasing irradiation fluence and decreasing temperature.
% 	\label{fig:torque_14T}
% 	}
% \end{figure*}

\clearpage

\section{Determination of the specific-heat peak temperature}
Figure\,S5 presents the procedure used to determine the characteristic peak temperature in the specific heat $T_\mathrm{peak}$. As shown in Figs.\,S5(b) to S5(d), the 14\,T data are subtracted as a background, and the peak positions are identified as indicated by the arrows. While the peak temperature $T_\mathrm{peak}$ is nearly independent of $H$ for the pristine sample, $T_\mathrm{peak}$ shifts systematically to higher temperatures for irradiated samples [Fig.\,S5(a)].

\begin{figure*}[h]
	\includegraphics[clip,width=12cm]{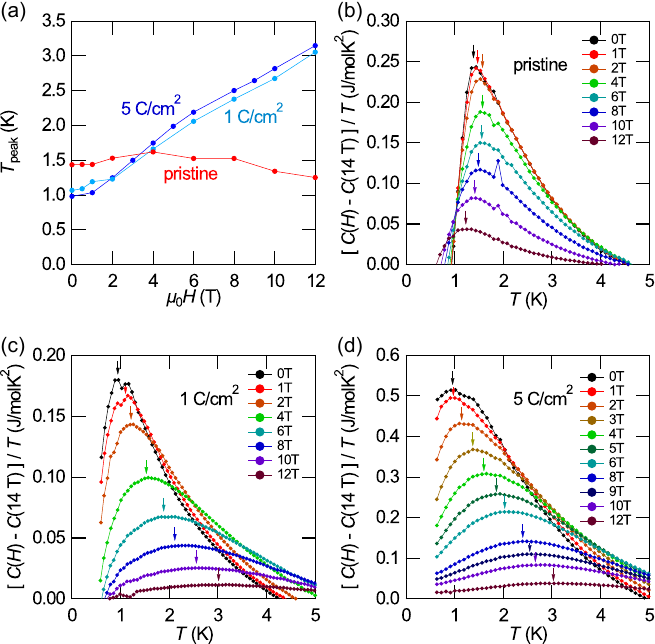}
	\caption{Magnetic specific heat of pristine and irradiated YCOB. (a) Field dependence of the peak temperature $T_\mathrm{peak}$ for pristine, 1\,C/cm$^2$, and 5\,C/cm$^2$ samples, respectively. (b-d) Temperature dependence of $[C(H) - C(14\ \mathrm{T})] / T$ for pristine (b), 1\,C/cm$^2$ (c), and 5\,C/cm$^2$ (d) samples, respectively.
	}
\end{figure*}

\section{Scaling analysis without subtracting the phonon contribution}
For comparison, Fig.\,S6 shows the scaling plots constructed from the total measured specific heat $C$, without subtracting the phonon contribution. These plots display $(\mu_0 H)^\gamma C/T$ as a function of $T/H$ for the pristine, 1\,C/cm$^2$, and 5\,C/cm$^2$ samples, using the same values of $\gamma$ as those employed in Fig.\,5 of the main text. At large $T/H$, the scaling behavior is affected by the increasing phonon contribution, which does not obey the $T/H$ scaling. In contrast, in the low-temperature region, where the phonon contribution is sufficiently small, the scaling plots constructed from the raw specific heat exhibit the same qualitative trends as those obtained from $C_{\mathrm{mag}}=C-C_{\mathrm{ph}}$. In particular, the pristine sample shows no clear scaling collapse, whereas the collapse becomes progressively more pronounced with increasing irradiation fluence. These results demonstrate that the contrasting low-temperature scaling behavior between the pristine and irradiated samples is not an artifact of the subtraction of $C_\mathrm{ph}$.

\begin{figure*}[h]
	\includegraphics[clip,width=16cm]{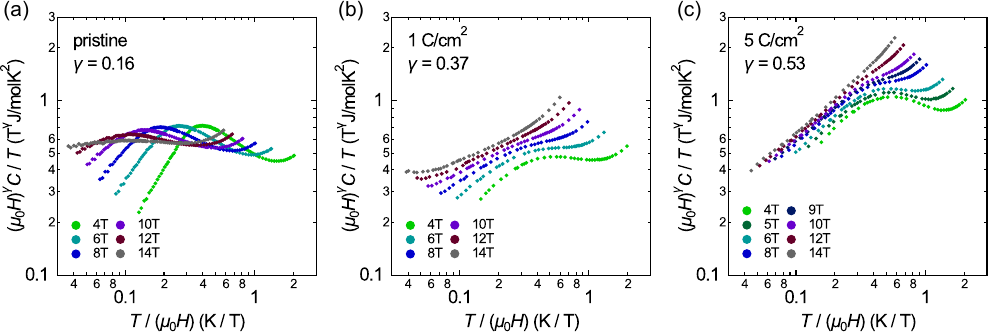}
	\caption{Scaling analysis of the specific heat without subtracting the phonon contribution. (a-c) Scaling plots of the total measured specific heat $(\mu_0 H)^\gamma C/T$ as a function of $T/H$ for the pristine (a), 1\,C/cm$^2$ (b), and 5\,C/cm$^2$ (c) samples. The phonon contribution has not been subtracted, and the values of $\gamma$ are the same as those used in Fig.\,5 of the main text. Deviations from scaling become evident at high temperatures owing to the increasing phonon contribution, whereas the low-temperature data exhibit the same qualitative trends as the scaling plots of $C_{\mathrm{mag}}$.}
	\label{fig:C_scaling_before_subtraction}
\end{figure*}

\bibliography{ref_supp.bib}